# Magnetic structures and phase transitions of A-site Ordered Double-Perovskite SmBaMn$_2$O$_6$

R. Kiyanagi[1,*], S. Yamada[2], H. Aoki[2], H. Sagayama[3], T. Moyoshi[4,6], A. Nakao[4], T. Arima[5]

[1]J-PARC center, Japan Atomic Energy Agency, Tokai, Ibaraki 319-1195, Japan
[2]Department of Life Science, Yokohama City University, Yokohama, Kanagawa 236-0027, Japan
[3]Institute of Materials Structure Science, High Energy Accelerator Research Organization, Tsukuba, Ibaraki 305-0801, Japan
[4]Comprehensive Research Organization for Science and Society (CROSS), Tokai, Ibaraki 319-1106, Japan
[5]Department of Advanced Materials Science, University of Tokyo, Kashiwa, Chiba 277-8561, Japan
[6]Deceased
* ryoji.kiyanagi@j-parc.jp

**ABSTRACT**

Magnetic structures and the relationship between spin and charge-orbital orderings of an A-site ordered double-perovskite manganite SmBaMn$_2$O$_6$, an anticipated multiferroic material, were investigated by means of neutron diffraction. The spin arrangement in MnO$_2$ planes perpendicular to the *c* axis is revealed to be the same as that in the A-site disordered half-doped manganites CE-type but the stacking pattern is found to be different displaying a unique twofold period. The temperature dependence of the superlattice magnetic and nuclear reflections clarifies that the antiferromagnetic spin ordering occurs at a temperature slightly lower than the temperature at which a rearrangement of the charge-orbital orderings occurs. The result evidences that the rearrangement leads the spin ordering. The intensities of the magnetic reflections are found to change across $T_f$ = 10 K, suggesting a spin-flop by 90˚ while keeping the Mn spin ordering pattern unchanged.

**Introduction**

Perovskite-type manganese oxides $RE_{1-x}AE_xMnO_3$ (RE = trivalent rare earth, AE = divalent alkaline earth) are one of the most extensively studied materials because of their intriguing physical properties such as the colossal magnetoresistance effect and multiferroicity.[1-6] In this system, charge, orbital and spin degrees of freedom strongly correlate with each other and exhibit a variety of electric phases depending on the combination of the elements RE and AE. While, in general, RE and AE randomly occupy the A-sites in the perovskite-type structure, in the case of AE = Ba and $x = 0.5$ it has been discovered that RE and Ba ions can alternately occupy the A-sites along the $c$ axis, as seen in Fig. 1, if the sample preparation conditions are properly controled.[7-9] Since this discovery, the A-site ordered $REBaMn_2O_6$ has also been extensively studied, and a variety of physical phenomena, as in the A-site disordered materials, has been showcased.[10-25]

A theoretical study recently predicted existence of spontaneous electric polarization in $SmBaMn_2O_6$, and the material has drawn attention from the view point of multiferroicity. $SmBaMn_2O_6$ shows two-step charge-orbital orderings (COO) and an antiferromagnetic (AFM) transition as the temperature decreases. The COOs take place around 380 K ($T_{CO1}$) and 200 K ($T_{CO2}$). The COO patterns in $T_{CO1} > T > T_{CO2}$ (HT-COO phase) and in $T < T_{CO2}$ (LT-COO phase) have already been elucidated to be similar to the one commonly seen in the A-site disordered half-doped manganites.[22] The phase transition from HT-COO to LT-

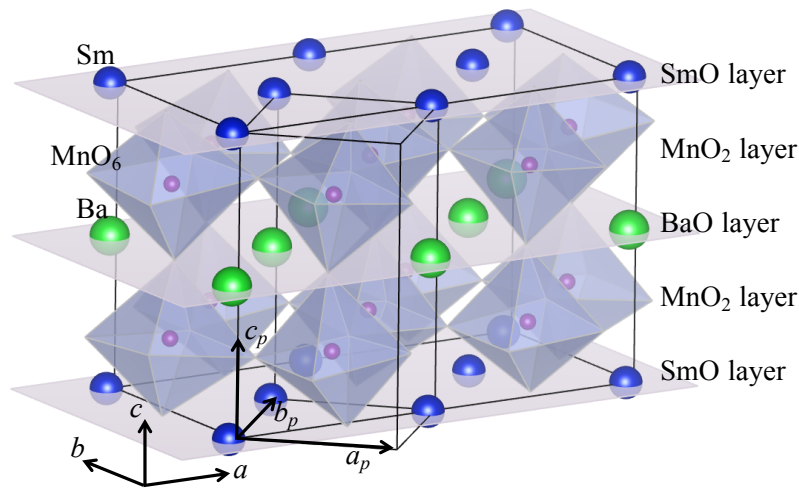

Figure 1. Schematic drawing of the crystal structure of $SmBaMn_2O_6$. The arrows with $a \times b \times c$ denote the crystallographic unit cell and the ones with $a_p \times b_p \times c_p$ denote a pseudo-tetragonal primitive unit cell ($a \times b \times c = 2\sqrt{2}a_p \times \sqrt{2}b_p \times c_p$). The notation used in this paper are based on the cell of $\sqrt{2}a_p \times \sqrt{2}b_p \times c_p$ (= $a/2 \times b \times c$). The structure is drawn by the program VESTA[28].

COO phases involves structural changes; namely the change of the space group from nonpolar *Pnam* to polar *P2₁am*.[21,25] A precise structural study by synchrotron diffraction revealed shift of oxygen ions in $MnO_2$ layers from the regular positions in the LT-COO phase, inducing local and spontaneous electronic polarization as predicted.[23-25]

As for magnetic properties, susceptibility measurements demonstrated an AFM phase transition at $T_N$ = 175 K followed by an abrupt change of the anisotropy at $T_f$ = 10 K.[22] Meanwhile the magnetic structure of $SmBaMn_2O_6$ has not been studied so far. This is because Sm has a huge neutron absorption cross section, which makes neutron experiments on this material very challenging. Nonetheless the magnetic structure is indispensable information for understanding the nature of multiferroicity in $SmBaMn_2O_6$. In addition, Yamada *et al.* recently discovered that the behaviors of the COO and the magnetic properties in $SmBaMn_2O_6$ strongly depend on the form of the measured samples, namely pulverized or a single crystal.[22] The magnetic susceptibility showed a very broad peak, ranging from 300 K to 100 K, for pulverized samples, while single crystal samples showed much sharper anomalies at $T_N$ and $T_f$. These observations clearly illustrate the importance of the use of single crystals for the detailed study of $SmBaMn_2O_6$. In this paper, we report the first study of the magnetic structure and the relationship between the magnetic ordering and the COO of $SmBaMn_2O_6$ by means of single-crystal neutron diffraction.

**Results and Discussion**

Despite the huge neutron absorption cross section of Sm, a large reciprocal space was clearly measured as exemplified in Fig. 2. In the reciprocal map on the $l$ = 0 plane at 100 K, many Bragg reflections can be clearly recognized (Fig. 2(a)). Here, the reciprocal lattice is drawn based on the pseudo-tetragonal cell of $\sqrt{2}a_p \times \sqrt{2}b_p \times c_p$ (see Fig. 1). Some reflections are found at ($h$+1/2 $k$ 0) as pointed by red arrows in Fig. 2(a), and disappearance of the superlattice reflections with a propagation vector of $q_l$ = (0 0 1/2) were confirmed at 100 K (the inset of Fig. 2(b)). These observations are consistent with the preceding studies, proposing the unit cell of $2\sqrt{2}a_p \times \sqrt{2}a_p \times c_p$ in the LT-COO phase. Note that there are also superlattice reflections at ($h$ $k$+1/2 0) as indicated by blue arrows in Fig. 2(a). These reflections can be ascribed to the other twinned domain in the sample.

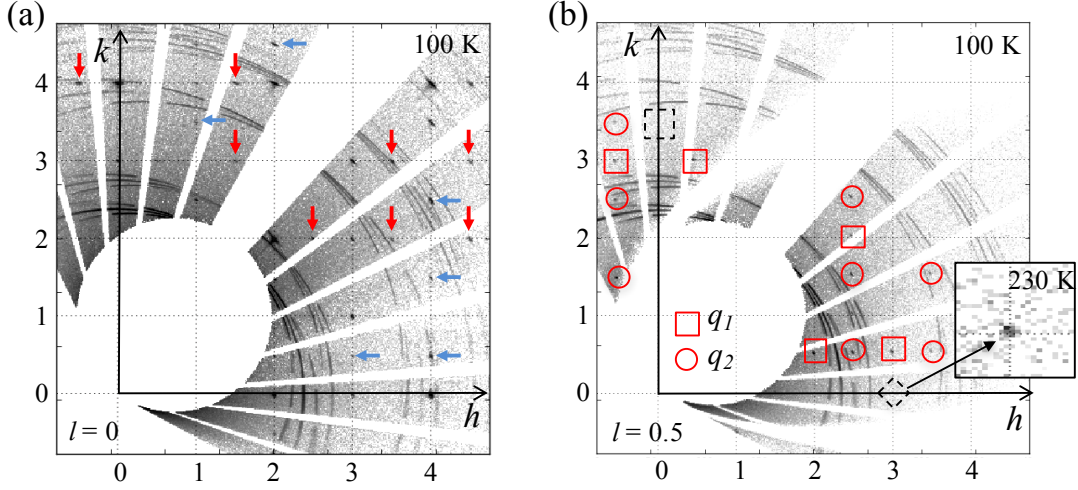

Figure 2. (a) Observed reciprocal maps on $l = 0$ plane at 100 K. The red arrows indicate ($h$+1/2 $k$ 0) superlattice reflections corresponding to the cell doubling along the $a$ axis. The blue ones are ($h$ $k$+1/2 0) superlattice reflections from the other twinned domain. (b) Observed reciprocal map on $l = 0.5$ plane at 100 K. The black broken diamond on the $a^*$ axis indicates the disappearance of the superlattice reflection of $q_1$, which is seen at 230 K. The black broken square on the $b^*$ axis highlights the absence of (0 3.5 0.5) reflection (See Fig. 5(a) for comparison).

Superlattice reflections with propagation vectors of $q_1$ = (0 1/2 1/2), (1/2 0 1/2) and $q_2$ = (1/2 1/2 1/2) were found at 100 K as shown in Fig. 2(b). None of these new reflections has been reported in the previous X-ray nor electron diffraction measurements, and the manifestations of the reflections, as is mentioned later, are in accord with the behavior of the magnetic susceptibility. Additionally, these reflections were found only in a small $Q$ region. For these reasons, it should be reasonable to assume that these reflections arise from the ordering of the magnetic moments, namely the spins of $Mn^{3+}$ and $Mn^{4+}$ ions. It should be mentioned that the propagation vector $q_1$ cannot be uniquely determined merely from the observed data since the sample contains twined domains sharing the $c$ axis. According to the observed reflection pattern, there are two candidates as the magnetic propagation vectors, (1/2 0 1/2) and (0 1/2 1/2). Since the propagation vector of the orbital ordering in the LT-COO phase has already been clarified to be (1/2 0 0), the propagation vectors of the magnetic ordering should be (0 1/2 $l$) and (1/2 1/2 $l$). We hence concluded that the magnetic order has the propagation vectors $q_1$ = (0 1/2 1/2) and $q_2$ = (1/2 1/2 1/2). The observed propagation vectors are identical with the ones observed in $YBaMn_2O_6$ of which in-plane components are the ones frequently seen in the A-site disordered half-doped manganites, so-called CE-type.[16] The propagation vectors, $q_1$ and $q_2$, represent the periodicities of the $Mn^{3+}$ and $Mn^{4+}$ magnetic moments, respectively. That is, the magnetic moments of $Mn^{3+}$ align ferro magnetically along the *a* axis and antiferromagnetically along the *b* axis. The magnetic moments of $Mn^{4+}$ are arranged in the staggered antiferromagnetic manner in the *a-b* plane. Drawn in Fig. 3(a) is the arrangement of the magnetic moments of $Mn^{3+}$ and $Mn^{4+}$ in the *a-b* plane superimposed on the expected COO pattern.

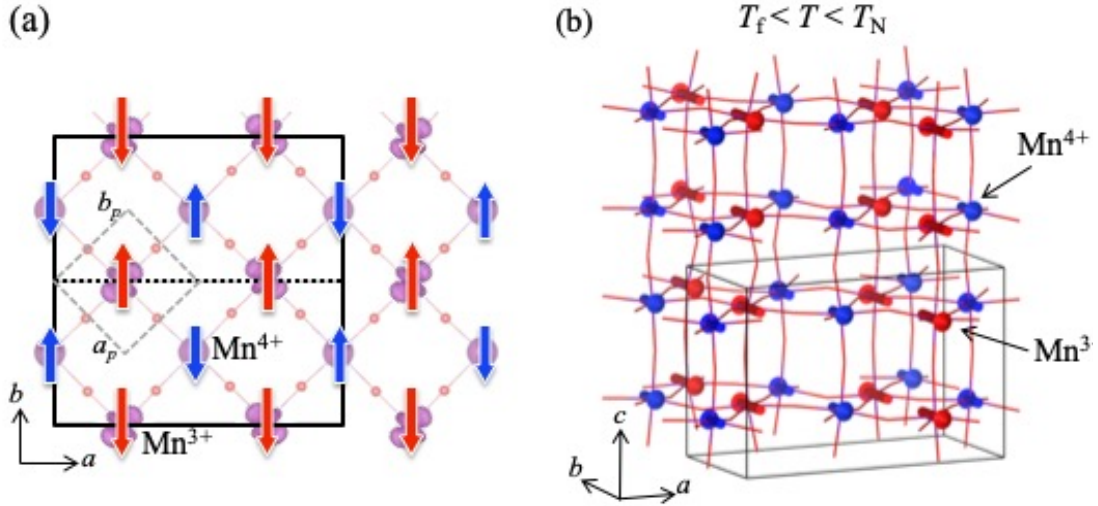

Figure 3. (a) Arrangement of the magnetic moments of the $Mn^{3+}$ and $Mn^{4+}$ on the *a-b* plane in $T_f < T < T_N$ together with the expected COO pattern in the LT-COO phase. The black solid lines correspond to the magnetic unit cell and the dotted line corresponds to the crystallographic unit cell. (b) Deduced 3-dimensional configuration of the magnetic moments in $T_f < T < T_N$ (The magnetic structure can be shifted by a quarter with respect to the magnetic unitcell).

Since the periodicity along the *c* axis is twofold, the stacking pattern should be of an AABB-type, where the coupling between A and B is antiferromagnetic. Since Sm and Ba ions alternately reside at the A-sites along the *c* axis, the Mn-Mn coupling along the *c* axis is ferromagnetic either across the SmO or BaO layer and antiferromagnetic across the other layer. According to the synchrotron structure analyses, the Mn-O-Mn angles across the SmO layer and the BaO layer noticeably differ from each other. This might be the cause of the different couplings, anti- or ferromagnetic, across the SmO or BaO layers. The stacking manner of the magnetic ordering along the *c* axis is a clear contrast to the one of the COO. The COO in the LT-COO phase shows in-phase stacking along the *c* axis, as in the A-site disordered half-doped manganites. Note that the COO in the HT-COO phase is twofold along the *c* axis.

The direction of the magnetic moments can be deduced to parallel to the *b* axis because there are no strong magnetic reflections observed on the *b\** axis. Magnetic reflections on the $b^*$ axis must be strong if the magnetic moments were parallel to the $a$ axis. The overall ordering pattern of the magnetic moments of the $Mn^{3+}$ and $Mn^{4+}$ in $T_f < T < T_N$ is depicted in Fig. 3(b).

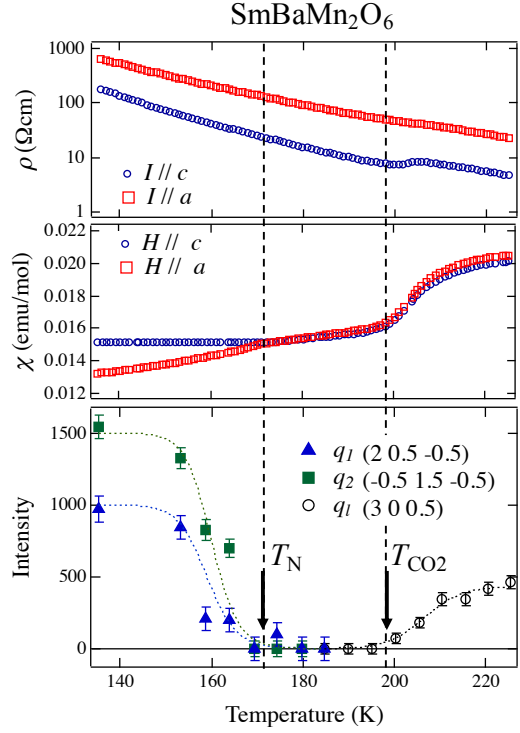

Figure 4. Variation of resistivity, susceptibility and the intensities of the magnetic reflections, $q_1$ and $q_2$, and the superlattice reflection, $q_l$, with heating. The resistivity and susceptibility data are excerpts from Ref. 22.

The detailed relationship between the rearrangement of COO at $T_{CO2}$ and the antiferromagnetic ordering at $T_N$ was examined by tracking the temperature dependence of the magnetic reflections at $q_1$ and $q_2$ and the superlattice reflections at $q_l$ that represents the HT-COO phase. The variations of the reflection intensities are shown in Fig. 4 together with the resistivity and magnetic susceptibility data. All the data were measured in a heating process. The intensities of the two magnetic reflections, $q_1$ and $q_2$, decrease as the temperature increases and disappear around 170 K, which well coincides with the temperature at which the anisotropy in the susceptibility disappears. The superlattice reflection at $q_l$ starts to emerge around 200 K. Since there is a clear gap between the temperatures at which the magnetic reflections disappear and the $q_l$ reflection starts to appear, it can be conjectured that the COO rearrangement into the LT-COO phase and the AFM phase transition do not simultaneously occur but the LT-COO phase leads the AFM phase transition.

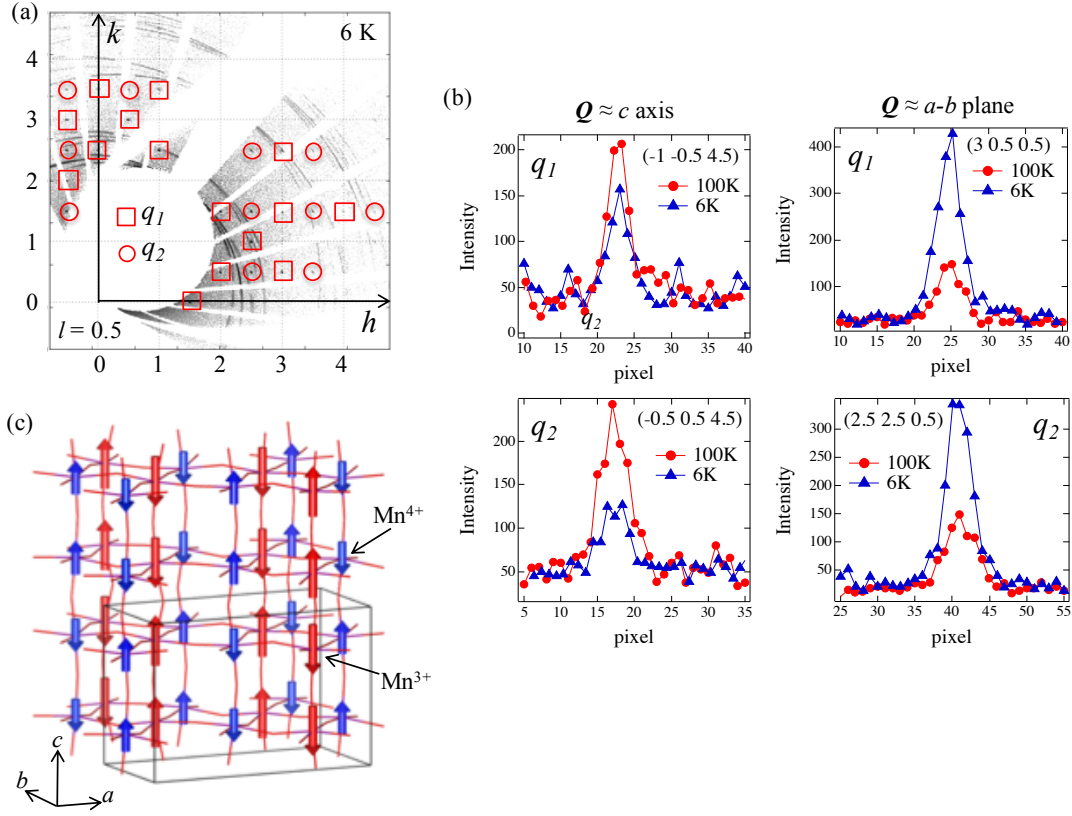

Figure 5. (a) Observed reciprocal map on $l = 0.5$ plane at 6 K. (b) Comparison of some magnetic reflections between 100 K and 6 K. (c) Expected 3-dimensional configuration of the magnetic moments of the $Mn^{3+}$ and $Mn^{4+}$ in $T < T_f$ (The magnetic structure can be shifted by a quarter with respect to the magnetic unitcell).

The data at 6 K, below $T_f$, show the similar diffraction pattern to 100 K, as shown in Fig. 5(a). The locations of the magnetic reflections, $q_1$ and $q_2$, are found to be unchanged and no new reflections were observed, indicating that the spin configuration remains the same. However, a significant change in the intensities of the magnetic reflections is noticed. Shown in Fig. 5(b) is the comparisons of the magnetic reflections between 100 K and 6 K. The intensities of the reflections with large $l$ components are weakened at 6 K while the reflections with large $h$ and small $l$ are enhanced. The same tendencies were found for both $q_1$ and $q_2$ reflections. Therefore, the magnetic moments of both $Mn^{3+}$ and $Mn^{4+}$ are considered to flop by 90° to the $c$ axis. The flop has been inferred from the magnetic susceptibility measurement as well. The magnetic susceptibility of $SmBaMn_2O_6$ shows a sudden drop at $T_f$ when the magnetic field is applied parallel to the $c$ axis while showing a steep rise when the magnetic field is parallel to the *a* axis. The flop is presumably induced by the magnetic moment of Sm ions, yet further thorough quantitative studies including $Sm^{3+}$ magnetic moment are required to clarify the details. The deduced configuration of the magnetic moments of the $Mn^{3+}$ and $Mn^{4+}$ in $T < T_f$ is depicted in Fig. 5(c).

**Conclusions**

The magnetic structures and phase transitions of the A-site ordered double-perovskite-type material $SmBaMn_2O_6$ were investigated by means of neutron single-crystal diffraction. Despite the very strong neutron absorption of Sm, the neutron diffraction pattern clearly showed magnetic superlattice reflections. The observed patterns of the magnetic reflections indicated the CE-type spin ordering in the $MnO_2$ plane with a twofold stacking period along the *c* axis. Thorough temperature dependent measurements of the superlattice reflections revealed that the rearrangement of the COO leads, not directly coupled with, the antiferromagnetic spin ordering. The Mn magnetic moments were found to flop by 90° from the *b* axis to the *c* axis across $T_f$ while the spin ordering pattern remains unchanged. Although the flop is presumably induced by the magnetic moments of $Sm^{3+}$ ions, quantitative magnetic structure analyses should be required in order to clarify the comprehensive magnetic structure including $Sm^{3+}$ magnetic moments below $T_f$, which also sheds light on the relationship between the Mn-Mn anti-/ferromagnetic couplings along the *c* axis and the SmO/BaO layers.

**Methods**

A single crystal of $SmBaMn_2O_6$ was grown by the floating zone method. The details of the sample preparation are described in Ref. 22. The dimension of the single crystal used in the neutron diffraction experiment was about $2 \times 1.2 \times 4$ mm$^3$.

The neutron diffraction experiment was conducted on the time-of-flight neutron single-crystal diffractometer, SENJU, built in the Materials and Life science experimental facility (MLF) in J-PARC.[26] SENJU is designed for neutron single-crystal structure analyses mainly targeting inorganic materials as well as magnetic materials. SENJU is equipped with 37 two-dimensional detectors around the sample position at the center of the instrument. Neutrons with a wide range of wavelengths are available on SENJU, which, together with many detectors, makes the diffraction measurements highly efficient. The single crystal sample was mounted on a closed-cycle type $^4$He-gas refrigerator. The refrigerator has two rotation axes on the cold finger so as to rotate the sample without moving the refrigerator. The refrigerator was placed in a vacuum chamber in order to reduce the background noises and also for thermal insulation. Measurements were carried out at 230 K, 100 K and 6 K, corresponding to the HT-COO phase, LT-COO phase and below $T_f$, respectively. Temperature dependence of some superlattice reflections were also measured around $T_N \sim 200$ K in order to thoroughly examine the AFM and COO phase transitions. The software STARGazer was used to process and visualize the data [27].

**Acknowledgment**

The measurement was performed using SENJU at MLF in J-PARC along with the project use program (2014P0906) of J-PARC. This work was partly supported by JSPS KAKENHI Grant Number JP24540380 and JP1818K03546

**Author contributions**

S.Y., H.S. and T.A. conceived of the presented idea. S.Y. and H.A. prepared the sample used in the experiment. R.K., S.Y., H.A., H.S., T.M., A.N. and T.A. conducted the experiment. R.K. analyzed the data, and R.K., S.Y, H.S. and T.A. discussed the results. All authors contributed to the final manuscript.

**Competing Interests**

The authors declare no competing interests.

**Data availability statement**